\documentclass[10pt]{article}
\newcommand{\cleqn}{\setcounter{equation}{0}}

\usepackage{graphicx,epsfig,citesort}
\usepackage{latexsym}

\begin{document}

\begin{flushright}
SU-4252-752\\
UM-P 017-2002\\
RCHEP 002-2002\\
hep-th/0207043
\end{flushright}

\begin{center}
{\Large\bf Non-Abelian Monopole and Dyon Solutions in a Modified
Einstein-Yang-Mills-Higgs System.}\\
\hspace{10pt}\\

A.S. Cornell$^{a,}$\footnote{\tt a.cornell@tauon.ph.unimelb.edu.au},
G.C. Joshi$^{a,}$\footnote{\tt joshi@physics.unimelb.edu.au},
J.S. Rozowsky$^{b,}$\footnote{\tt rozowsky@phy.syr.edu},
K.C. Wali$^{b,}$\footnote{\tt wali@physics.syr.edu}\\
\vskip 0.2cm
{$^a$\em Research Centre for High Energy Physics, School of Physics,}\\
{\em University of Melbourne, Parkville, Victoria 3010, Australia.}\\
{$^b$\em Physics Department, Syracuse University, Syracuse, NY 13244.}\\

\vskip .2cm
\today
\end{center}

\begin{abstract}
We have studied a modified Yang-Mills-Higgs system coupled to 
Einstein gravity. The modification of the Einstein-Hilbert 
action involves a direct coupling of the Higgs field to
the scalar curvature. In this modified system we are able to
write a Bogomol'nyi type condition in curved space and demonstrate 
that the positive static energy functional is bounded from below.
We then investigate non-Abelian spherically symmetric static 
solutions in a similar fashion to the 't Hooft-Polyakov monopole.
After reviewing previously studied monopole solutions of this 
type, we extend the formalism to included electric charge and 
we present dyon solutions. 
\end{abstract}

\section{Introduction}
In a classic paper \cite{Dirac}, Dirac proposed the
possible existence of a magnetic monopole, the analogue of an
isolated electrically charged particle.
 Motivated principally to restore the symmetries between electric and magnetic
forces, Dirac found that the existence of a monopole provided a
natural explanation for the quantization of electric charge.  A
description of such a monopole consistent with quantum mechanics
would lead to the famous Dirac charge quantization condition,
\begin{eqnarray}
eg = \frac{1}{2}n\hbar c, \nonumber
\end{eqnarray}
where $e^2/\hbar c$ is the fine structure constant, $g$ is the
monopole charge, and $n$ an integer.  Dirac's theory required a
$U(1)$ valued gauge potential which was singular along a line
(Dirac string) originating from the monopole and extending to
infinity.  Later Dirac's theory was reformulated by Wu and 
Yang~\cite{Wu} within the framework of fiber bundles.  The singular
line is avoided at the expense of introducing  coordinate patches
on a sphere surrounding the monopole.
 However, the transition functions between the coordinate patches that are
elements of the $U(1)$ gauge group are singular. Thus, there is a
complete equivalence between the two descriptions~\cite{Primac} and
the monopole emerges on a sound footing like any other particle in
nature. Consequently, during the past decades, extensive effort
has gone into experimental search for monopoles, but unfortunately 
has as yet had no success.

In spite of the lack of experimental
success, the monopole continues to thrive in the theoretical
laboratory. With the pioneering work of 't Hooft and Polyakov
\cite{Polyakov}, the monopole was reinvented in a new form as a finite
energy, particle-like soliton in non-Abelian gauge theories with
spontaneous symmetry breaking.
 Moreover, such objects are generic in any spontaneously broken
non-Abelian gauge theory which has an unbroken $U(1)$ gauge
symmetry. Such monopoles are expected to be produced
in abundance in phase transitions of grand unified theories, which 
has implications for early universe cosmology. New searches
for such relic monopoles are under way \cite{search,signatures}.

More recently, a great deal of activity has centered around monopoles in
curved space-time in order to study the effects of gravity. New 
insights have emerged from a study of a coupled Einstein-Yang-Mills-Higgs
[EYMH] system with solutions describing black holes with magnetic
charge, black holes within magnetic monopoles and magnetic
monopoles within black holes~\cite{multi,Volkov,Lue,Brihay,Schap}.
Consequently, gravity cannot be dispensed with, arguing that the
strength of its interaction is weak.  The coupled set of equations
for the EYMH system lead to non-trivial consequences and provide a
fertile ground for the study of the interplay between gravitation
and other interactions.

The present paper is an extension of the work by Nguyen and 
Wali~\cite{Nguyen} to include electric charge and to study
dyons coupled to gravity. The starting point is a modified EYMH
system with a specific coupling of the Higgs field to the Einstein
term in the action. In the static case, this enables us, with the
help of a Bogomol'nyi-type~[9] condition, to reduce the energy
functional to a form that resembles the energy functional in flat
space-time and derive a lower bound on the energy and
hence the mass of the dyon in terms of its electric and magnetic charges.

In the next section, we begin with a review of the general
formalism and the field equations for the coupled EYMH system.  In
Section 3, we derive the field equations and the energy functional
in the static case. Through the Higgs field equation, we express
the Einstein term in the action in terms of metric fields, find a
positive definite expression for the energy functional, and derive
a lower bound on the energy.  We also discuss in this section, the
relation between the mass and the charge of the dyon. Section 4
is devoted to the derivation of the basic equations in the context
of a spherically symmetric static metric and the spherically
symmetric 't Hooft-Polyakov ansatz for the gauge and Higgs fields.
In Section 5, we specialize to the Higgs vacuum and find monopole and
dyon solutions. The final section is devoted to some concluding 
remarks.

\section{General Framework; Field Equations}
We begin by defining the action, 
\begin{equation}\label{21}
S = \int\,d^4x \sqrt{-g}\left( {\cal L}_E + {\cal L}_M \right) ,
\end{equation}
where
\begin{equation}\label{22}
{\cal L}_E = -\frac{R-2\Lambda}{16\pi G v^2} \mathbf{\Phi}^2 ,
\end{equation}
with $R$, the Ricci scalar, $\Lambda$, the cosmological constant 
and $\mathbf{\Phi}$, the Higgs scalar field. Our metric $g_{\mu\nu}$ 
is chosen to have signature $(+---)$ with indicies 
$\mu,\nu,\ldots$ running from 0 to 3 and indicies $i,j,\ldots$ 
from 1 to 3, also $g=\det |g_{\mu\nu}|$. The matter content is
given by
\begin{equation}\label{23}
{\cal L}_M = -\frac{1}{4}g^{\mu\rho}g^{\nu\lambda}
\mathbf{F}_{\mu\nu} \mathbf{F}_{\rho\lambda} +
\frac{1}{2}g^{\mu\nu} \left({\cal D}_{\mu} \mathbf{\Phi} \right)
\left(\mathbf{{\cal D}}_{\nu} \mathbf{\Phi} \right) -
\frac{\lambda}{4} \left( \mathbf{\Phi}^2 - v^2 \right)^2,
\end{equation}
where
\begin{equation}\label{24}
\mathbf{F}_{\mu\nu} = -\frac{1}{i\alpha}\left[ {\cal D}_{\mu}, {\cal D}_{\nu}
\right] ,
\end{equation}
with
\begin{equation}\label{25}
{\cal D}_{\mu} = \nabla_{\mu} - i\alpha [ \mathbf{A}_{\mu}, \cdot ] .
\end{equation}

Thus $\mathbf{F}_{\mu\nu}$ is the field strength associated with the
gauge field $\mathbf{A}_{\mu}$, $\alpha$ being the strength of the gauge
coupling.  $\nabla_{\mu}$ is the covariant derivative with the metric
compatible, torsion free connection coefficients.

More explicitly,
\begin{equation}\label{26}
\mathbf{F}_{\mu\nu} = \partial_{\mu} \mathbf{A}_{\nu} - \partial_{\nu}
\mathbf{A}_{\mu} - i\alpha [ \mathbf{A}_{\mu}, \mathbf{A}_{\nu} ] ,
\end{equation}
and in the component form,
\begin{equation}\label{27}
F^a_{\mu\nu} = \partial_{\mu} A^a_{\nu} - \partial_{\nu} A^a_{\mu} +
\alpha f^{abc} A^b_{\mu} A^c_{\nu} ,
\end{equation}
where $f^{abc}$ are the structure constants of a gauge group ${\cal G}$, which
for the most part for our purposes will be $SU(2)$.  The scalar field
$\mathbf{\Phi}$ belongs to the adjoint representation of ${\cal G}$.  Its
covariant derivative is given by
\begin{eqnarray}\label{28}
{\cal D}_{\mu} \mathbf{\Phi} & = & \nabla_{\mu} \mathbf{\Phi} - i\alpha
[ \mathbf{A}_{\mu}, \mathbf{\Phi} ] \nonumber \\
& = & \partial_{\mu} \mathbf{\Phi} - i\alpha [ \mathbf{A}_{\mu}, 
\mathbf{\Phi} ],
\end{eqnarray}
and in component form
\begin{equation}\label{29}
{\cal D}_{\mu} \Phi^a = \partial_{\mu} \Phi^a + \alpha f^{abc} A^b_{\mu} \Phi^c .
\end{equation}

We note that in the broken phase of the gauge symmetry, when the Higgs
field $\mathbf{\Phi}$ assumes its vacuum expectation value, $\mathbf{\Phi}^2 =
v^2$, ${\cal L}_E$ in (\ref{22}) is the conventional Einstein-Hilbert Lagrangian.
${\cal L}_M$ in (\ref{23}) represents the standard Yang-Mills-Higgs Lagrangian in
curved space-time.

By varying the action $S$ with respect to $\mathbf{A}_{\mu}$,
$\mathbf{\Phi}$ and $g_{\mu\nu}$, we obtain the coupled Yang-Mills, 
Higgs and Einstein equations of motion:
\begin{eqnarray}
\frac{1}{\sqrt{-g}} {\cal D}_{\mu} 
\left( \sqrt{-g} \mathbf{F}^{\mu\nu} \right) 
&=& i\alpha [\mathbf{\Phi}, {\cal D}^{\nu}\mathbf{\Phi} ] , \label{ym}\\
\frac{1}{\sqrt{-g}} {\cal D}_{\mu} 
\left( \sqrt{-g} {\cal D}^{\mu} \mathbf{\Phi}\right) 
&=& \left( \frac{R-2\Lambda}{8\pi Gv^2} + \lambda(\mathbf{\Phi}^2 -
v^2) \right)\mathbf{\Phi} , \label{higgs}
\end{eqnarray}
and
\begin{equation}
G_{\mu\nu} = R_{\mu\nu} - \frac{R-2\Lambda}{2}g_{\mu\nu}
=\frac{8\pi Gv^2}{\mathbf{\Phi}^2} T_{\mu\nu} \label{einstein},
\end{equation}
where the energy-momentum tensor $T_{\mu\nu}$ is given by
\begin{eqnarray}\label{213}
T_{\mu\nu} & = & -\left({\cal L}_M + \frac{1}{2}\Box
\mathbf{\Phi}^2\right) g_{\mu\nu} - \mathbf{F}_{\mu\rho}\cdot
\mathbf{F}_{\nu}^{\rho} + {\cal D}_{\mu}\mathbf{\Phi}\cdot {\cal
D}_{\nu}\mathbf{\Phi} + \nabla_{\mu} \nabla_{\nu} \mathbf{\Phi}^2.
\end{eqnarray}

The terms involving $\mathbf{\Phi}$ in  (\ref{213}) arise
because of its presence in the modified Einstein-Hilbert 
action~\cite{Parker} in eqn.~(\ref{22}). We further note that
although the field equations are written in terms of the covariant
derivative $\nabla_{\mu}$ in ${\cal D}_{\mu}$, we can easily show
that they can all be reduced to ordinary partial derivatives since
\begin{equation}
\nabla_{\sigma}g_{\mu\nu} = 0 \hspace{20pt} ; \hspace{20pt} \nabla_{\mu}
\sqrt{-g} = \partial_{\mu}\sqrt{-g} - \sqrt{-g}\Gamma^{\nu}_{\mu\nu} = 0,
\end{equation}
for torsion-free, metric compatible connection coefficients
$\Gamma^{\lambda}_{\mu\nu} = \Gamma^{\lambda}_{\nu\mu}$.
Therefore, henceforth, in our equations,
\begin{equation}
 \mathbf{D}_{\mu} =
\partial_{\mu}\mathbf{1} - i\alpha [ \mathbf{A}_{\mu} , \cdot ].
\end{equation}

\section{Static Field Equations; Bogomol'nyi Bound}
\cleqn 
We are interested in static solutions to 
equations (\ref{ym}) - (\ref{einstein}).  Setting the time derivatives
of all fields equal to zero, we find equation (\ref{ym}) reduces
to two equations,
\begin{eqnarray}
\frac{1}{\sqrt{-g}} \mathbf{D}_{i} \left(\sqrt{-g} \mathbf{F}^{i0} \right) & = &
i\alpha \left[ \mathbf{\Phi}, \mathbf{D}^0 \mathbf{\Phi} \right] , \label{31} \\
\mathbf{D}_{0} \mathbf{F}^{0j} + \frac{1}{\sqrt{-g}} \mathbf{D}_{i}
\left(\sqrt{-g} \mathbf{F}^{ij} \right) & = & i\alpha \left[ \mathbf{\Phi},
\mathbf{D}^j \mathbf{\Phi} \right] . \label{32}
\end{eqnarray}
The Higgs field equation (\ref{higgs}) becomes
\begin{eqnarray}\label{33}
\frac{1}{\sqrt{-g}} \left[ \mathbf{D}_{0} \left( \sqrt{-g} \mathbf{D}^0
\mathbf{\Phi} \right) + \mathbf{D}_{i} \left( \sqrt{-g} \mathbf{D}^i
\mathbf{\Phi} \right) \right] = - \left( \frac{R-2\Lambda}{8\pi G v^2} + \lambda
\left( \mathbf{\Phi}^2 - v^2 \right) \right) \mathbf{\Phi}.
\end{eqnarray}
In order to find a Bogomol'nyi-type first-order
equation~\cite{Bogo} to our problem, we make the
ansatz~\cite{Forgacs} ,
\begin{eqnarray}\label{34}
\mathbf{F}_{ij} & = & \sqrt{-\tilde{g}} \epsilon_{ijk} \left( \mathbf{D}^k + u^k
\right) \mathbf{\Phi} ,
\end{eqnarray}
where $u^k = \partial^k f$ is an arbitrary time-independent function, and
$\tilde{g} = \det | g_{ij} |$.

With the above ansatz, we find
\begin{eqnarray}\label{35}
\mathbf{D}_{i} \mathbf{F}^{ij} = i\alpha \left[ \mathbf{\Phi}, \mathbf{D}^j
\mathbf{\Phi} \right] + \left( \frac{\partial_i \sqrt{g_{00}} }{ \sqrt{g_{00}}} -
\partial_i f \right) \mathbf{F}^{ij} .
\end{eqnarray}
Substituting (\ref{35}) in (\ref{32}), we have
\begin{eqnarray}\label{36}
\left[ \mathbf{D}_{0} , \mathbf{F}^{0j} \right] + \left( \frac{\partial_i
\sqrt{g_{00}}}{ \sqrt{g_{00}}} - \partial_i f \right) \mathbf{F}^{ij} = 0.
\end{eqnarray}
For future reference, we note that the Yang-Mills equation (\ref{31}) implies
\begin{eqnarray}\label{37}
\mathbf{D}_{i} \left( \sqrt{-g} \mathbf{F}^{i0} \right) 
\cdot \mathbf{\Phi} = 0.
\end{eqnarray}
Using (\ref{34}) together with (\ref{35}), after some algebra 
and the use of the Bianchi identity, the Higgs field 
equation (\ref{33}) reduces to
\begin{eqnarray}\label{38}
\mathbf{D}_{0} \mathbf{D}^{0} \mathbf{\Phi} & + & \left( \frac{\partial_i
\sqrt{g_{00}}}{ \sqrt{g_{00}}} - \partial_i f \right) \mathbf{D}^i \mathbf{\Phi}
- \frac{1}{\sqrt{-\tilde{g}}} \partial_i \left( \sqrt{-\tilde{g}} \partial^i f
\right) \mathbf{\Phi} \nonumber \\
& = & - \left( \frac{R-2\Lambda}{8\pi G v^2} + \lambda
\left( \mathbf{\Phi}^2 - v^2 \right) \right) \mathbf{\Phi} .
\end{eqnarray}
In deriving the above equations, with static spherically symmetric solutions in
mind, we have assumed
\begin{eqnarray}\label{39}
g_{00} > 0 \hspace{20pt} g_{0i} = 0 , \hspace{20pt} {\rm and} \hspace{20pt} 
-\tilde{g} > 0.
\end{eqnarray}

The static energy functional, ${\cal E}$, that follows from (\ref{21}) is
given by
\begin{eqnarray}\label{310}
{\cal E} & = & \int \,d^3x \frac{\sqrt{-g}}{2} \left[ \frac{R-2\Lambda}{8\pi G
v^2}\Phi^2 - \mathbf{F}^{0i}\cdot \mathbf{F}_{0i} + \frac{1}{2} \mathbf{F}^{ij} \cdot
\mathbf{F}_{ij} - \mathbf{D}^i \mathbf{\Phi} \cdot \mathbf{D}_i
\mathbf{\Phi} +
\frac{\lambda}{2}\left( \mathbf{\Phi}^2 - v^2 \right)^2 \right] ,
\end{eqnarray}
Next we define electric and magnetic fields, $\mathbf{E}_i$, $\mathbf{B}_i$ 
and corresponding $\mathbf{E}^i$, $\mathbf{B}^i$ to be given by
\begin{eqnarray}
\mathbf{E}_i = \frac{\sqrt[4]{-g}}{\sqrt{g_{00}}} \mathbf{F}_{0i} & ; &
\mathbf{E}^i = \sqrt[4]{-g} \sqrt{g_{00}} \mathbf{F}^{i0} , \label{311} \\
\mathbf{B}_i = \frac{1}{2} \sqrt[4]{-g} \sqrt{-\tilde{g}}
\epsilon_{ijk} \mathbf{F}^{jk} & ; & \mathbf{B}^i = \frac{1}{2}
\frac{\sqrt[4]{-g}}{ \sqrt{-\tilde{g}}} \epsilon^{ijk}
\mathbf{F}_{jk} , \label{312}
\end{eqnarray}
and
\begin{eqnarray}\label{313}
\mathbf{\chi} = \sqrt{g_{00}} \mathbf{\Phi} ,
\end{eqnarray}
where all the fields are functions of spacial coordinates only.

With these definitions and after substituting in (\ref{310}) for
$(R-2\Lambda)/{8\pi Gv^2}$ from (\ref{38}), we obtain
\begin{eqnarray}\label{314}
{\cal E} & = & {1\over 2} \int \,d^3x \left\{
-\left[ \mathbf{E}^i \cdot \mathbf{E}_i +
\mathbf{B}^i \cdot \mathbf{B}_i + \left( \frac{\sqrt[4]{-g}}{\sqrt{g_{00}}}
\mathbf{D}^i \mathbf{\chi} \right) \left( \frac{\sqrt[4]{-g}}{\sqrt{g_{00}}}
\mathbf{D}_i \mathbf{\chi} \right) 
- 2 \frac{\sqrt{-g}}{g_{00}} \mathbf{D}_0 \mathbf{\chi}
\cdot \mathbf{D}^0 \mathbf{\chi}\right] \right.\nonumber \\
&& \left. +\frac{1}{2}
\frac{\sqrt{-g}}{g_{00}} \left( \frac{\partial^i\sqrt{g_{00}}}{\sqrt{g_{00}}} 
- \partial^i f \right)\partial_i
\mathbf{\chi}^2 - \frac{\lambda}{2} {\sqrt{-g}\over g_{00}^2} \left(
\mathbf{\chi}^4 - g_{00}^2 v^4 \right)
+\partial_i \left( \frac{\sqrt{-g}}{g_{00}}
\mathbf{\chi}^2\partial^i f \right) \right\} .
\end{eqnarray}

In the spherically symmetric case in which we are interested, with the
assumed signature for the metric, we note that
\begin{eqnarray}\label{315}
g_{ij} = 0 , i\neq j \hspace{20pt} {\rm and} \hspace{20pt} -g_{ii} \geq 0
\end{eqnarray}
Consequently, each of the terms in the first parenthesis in (\ref{314}) is
positive.  They are exact analogues of the corresponding terms in the case of
flat space-time.  The next two terms
vanish in the Higgs vacuum, $\mathbf{\chi}^2 = g_{00} v^2$.  Finally, the last
term is a total divergence, giving rise to a finite surface term.  Ignoring the
terms that vanish when $\mathbf{\chi}^2 = g_{00} v^2$, we have a reduced
positive-definite energy functional, ${\cal E}$, given by
\begin{eqnarray}\label{316}
{\cal E}  &=&  \frac{1}{2} \int \,d^3x \left[ 
\mathbf{E}^i \cdot \mathbf{E}_i +
\mathbf{B}^i \cdot \mathbf{B}_i + \left( \frac{\sqrt[4]{-g}}{\sqrt{g_{00}}}
\mathbf{D}^i \mathbf{\chi} \right) \left( \frac{\sqrt[4]{-g}}{\sqrt{g_{00}}}
\mathbf{D}_i \mathbf{\chi} \right) \right. \nonumber \\
& & \left. \hskip 2.5cm
+ 2 \frac{\sqrt{-g}}{g_{00}} \mathbf{D}_0 \mathbf{\chi} \cdot
\mathbf{D}^0 \mathbf{\chi} + \partial_i \left( \frac{\sqrt{-{g}}}{g_{00}}
\mathbf{\chi}^2 \partial^i f \right) \right] .
\end{eqnarray}
As in the case of flat space-time~\cite{Coleman}, we can write
\begin{eqnarray}\label{317}
{\cal E} & = & \frac{1}{2} \int \,d^3x \left\{ \left[
\left(\mathbf{E}^i -\sin\theta \frac{\sqrt[4]{-g}}{\sqrt{g_{00}}}
\mathbf{D}^i \mathbf{\chi} \right) \left( \mathbf{E}_i -\sin\theta
\frac{\sqrt[4]{-g}}{\sqrt{g_{00}}} \mathbf{D}_i \mathbf{\chi}
\right)\right.\right.\\
&& \left.+ \left( \mathbf{B}^i - \cos\theta
\frac{\sqrt[4]{-g}}{\sqrt{g_{00}}} \mathbf{D}^i \mathbf{\chi}
\right) \left( \mathbf{B}_i - \cos\theta
\frac{\sqrt[4]{-g}}{\sqrt{g_{00}}} \mathbf{D}_i
\mathbf{\chi} \right) \right] 
+ 2\left[ \sin\theta \frac{\sqrt[4]{-g}}{\sqrt{g_{00}}} \mathbf{E}^i \cdot
\mathbf{D}_i \mathbf{\chi} \right.\nonumber \\
&& \left.
+ \cos\theta \frac{\sqrt[4]{-g}}{\sqrt{g_{00}}}
\mathbf{B}^i \cdot \mathbf{D}_i \mathbf{\chi} 
\left. + \frac{\sqrt{-g}}{g_{00}} \mathbf{D}_0 \mathbf{\chi} \cdot
\mathbf{D}^0 \mathbf{\chi} + \frac{1}{2} \partial_i \left(
\frac{\sqrt{-g}}{g_{00}} \mathbf{\chi}^2 \partial^i f \right) \right]
\right\}.
\end{eqnarray}
Hence, ${\cal E}$, has a lower bound,
\begin{eqnarray}\label{318}
{\cal E} & \geq & \int \,d^3x \left[\sin\theta \frac{\sqrt[4]{-g}}{\sqrt{g_{00}}}
\mathbf{E}^i \cdot
\mathbf{D}_i \mathbf{\chi} + \cos\theta \frac{\sqrt[4]{-g}}{\sqrt{g_{00}}}
\mathbf{B}^i \cdot \mathbf{D}_i \mathbf{\chi} 
+ \frac{\sqrt{-g}}{g_{00}} \mathbf{D}_0 \mathbf{\chi} \cdot
\mathbf{D}^0 \mathbf{\chi} + \frac{1}{2} \partial_i \left(
\frac{\sqrt{-{g}}}{g_{00}} \mathbf{\chi}^2 \partial^i f \right) \right], \quad
\end{eqnarray}
reaching the lower bound when the Bogomol'nyi-type equations are satisfied, 
that is when
\begin{eqnarray}\label{319}
\mathbf{E}_i -\sin\theta \frac{\sqrt[4]{-g}}{\sqrt{g_{00}}} \mathbf{D}_i
\mathbf{\chi} = \mathbf{B}_i - \cos\theta \frac{\sqrt[4]{-g}}{\sqrt{g_{00}}}
\mathbf{D}_i \mathbf{\chi} = 0 .
\end{eqnarray}
Now from our definition for $\mathbf{B}_i$ in (\ref{312}),
\begin{eqnarray}
\frac{\sqrt[4]{-g}}{\sqrt{g_{00}}} \mathbf{B}^i= \frac{1}{2} \epsilon^{ijk}
\mathbf{F}_{jk} , \nonumber
\end{eqnarray}
and from the Bianchi identity, we have the conservation law,
\begin{eqnarray}\label{320}
\mathbf{D}_i \left( \frac{\sqrt[4]{-g}}{\sqrt{g_{00}}} 
\mathbf{B}^i \right) = 0 .
\end{eqnarray}
Consequently,
\begin{eqnarray}\label{321}
\int \,d^3x \frac{\sqrt[4]{-g}}{\sqrt{g_{00}}} \mathbf{B}^i \cdot \mathbf{D}_i
\mathbf{\chi} = \int \,d^3x \mathbf{D}_i \left(
\frac{\sqrt[4]{-g}}{\sqrt{g_{00}}} \mathbf{B}^i \cdot \mathbf{\chi} \right) .
\end{eqnarray}
Similarly,
\begin{eqnarray}
\frac{\sqrt[4]{-g}}{\sqrt{g_{00}}} \mathbf{E}^i = \mathbf{F}^{i0} ,
\end{eqnarray}
and
\begin{eqnarray}\label{322}
\mathbf{D}_i \left( \frac{\sqrt[4]{-g}}{\sqrt{g_{00}}} \mathbf{E}^i \cdot
\mathbf{\chi} \right) & = & \mathbf{\chi} \cdot \mathbf{D}_i \mathbf{F}^{i0} +
\mathbf{E}^i \cdot \frac{\sqrt[4]{-g}}{\sqrt{g_{00}}} \mathbf{D}_i \mathbf{\chi}
\nonumber \\
& = & \mathbf{E}^i \cdot \frac{\sqrt[4]{-g}}{\sqrt{g_{00}}} \mathbf{D}_i
\mathbf{\chi} ,
\end{eqnarray}
on account of (\ref{37}).

Hence,
\begin{eqnarray}\label{323}
\int \,d^3x \left( \mathbf{E}^i \cdot \frac{\sqrt[4]{-g}}{\sqrt{g_{00}}}
\mathbf{D}_i \mathbf{\chi} \right) = \int \,d^3x \mathbf{D}_i \left(
\frac{\sqrt[4]{-g}}{\sqrt{g_{00}}} \mathbf{E}^i \cdot \mathbf{\chi} \right) .
\end{eqnarray}
When the Bogomol'nyi equation (\ref{319}) is satisfied, 
the lower bound on the
energy functional is satisfied and we have,
\begin{eqnarray}\label{324}
{\cal E} & = & \sin\theta \int \,d^3x \mathbf{D}_i \left(
\frac{\sqrt[4]{-g}}{\sqrt{g_{00}}} \mathbf{E}^i \cdot \mathbf{\chi} \right) +
\cos\theta \int \,d^3x \mathbf{D}_i \left( \frac{\sqrt[4]{-g}}{\sqrt{g_{00}}}
\mathbf{B}_i \cdot \mathbf{\chi} \right) \nonumber \\
& & + \int \,d^3x \left[ \frac{\sqrt{-g}}{g_{00}}
\mathbf{D}_0 \mathbf{\chi} \cdot \mathbf{D}^0\mathbf{\chi}+
{1\over 2}\partial_i \left( \frac{\sqrt{-{g}}}{g_{00}}
\mathbf{\chi}^2 \partial^i f \right)\right] .
\end{eqnarray}
Now, if we have finite-energy configurations with finite extension and
asymptotically flat space-time,
\begin{eqnarray}\label{325}
D_{\mu} \mathbf{\Phi} = 0 & , & \mathbf{\Phi}^2 = v^2,
\end{eqnarray}
leading to the condition that the gauge potential $\mathbf{A}_{\mu}$ is given by
\begin{eqnarray}\label{326}
\mathbf{A}_{\mu} = \frac{i}{\alpha v^2} \left[ \mathbf{\Phi} , \partial_{\mu}
\mathbf{\Phi} \right] + \frac{1}{v} \mathbf{\Phi} W_{\mu} ,
\end{eqnarray}
where $W_{\mu}$ is an arbitrary Abelian field.  The field strength
$\mathbf{F}_{\mu\nu}$ corresponding to the above gauge potential is
\begin{eqnarray}
\mathbf{F}_{\mu\nu} = \frac{1}{v} \mathbf{\Phi} F_{\mu\nu} , \nonumber
\end{eqnarray}
where
\begin{eqnarray}\label{327}
F_{\mu\nu} = \frac{i}{\alpha v^3} \left[ \partial_{\mu} \mathbf{\Phi} ,
\partial_{\nu} \mathbf{\Phi} \right] \cdot \mathbf{\Phi} 
+ \partial_{\mu} W_{\nu}
- \partial_{\nu} W_{\mu} .
\end{eqnarray}
In the static case, asymptotically,
\begin{eqnarray}
\mathbf{A}_{0} = \mathbf{\Phi} \frac{W_0}{v}, \quad  
\mathbf{A}_{i} = {i\over \alpha v^2} 
\left[ \mathbf{\Phi} , \partial_{i} \mathbf{\Phi} \right] , \nonumber
\end{eqnarray}
and
\begin{eqnarray}\label{328}
\mathbf{F}_{i0} & = & \mathbf{\Phi} \frac{1}{v} \partial_{i} W_0 = - \mathbf{E}_i
, \nonumber \\
\mathbf{F}_{ij} & = & \frac{1}{2v} \mathbf{\Phi} \left\{ \frac{i}{\alpha v^3}
\left[ \partial_{i} \mathbf{\Phi} , \partial_{j} \mathbf{\Phi} \right] \cdot
\mathbf{\Phi} \right\} ,
\end{eqnarray}
where we have assumed $W_i = 0$, $i=1,2,3$.

The above field configurations imply that outside a finite region, the
non-Abelian gauge field is purely in the direction $\mathbf{\Phi}$, the direction
of the unbroken $U(1)$ electromagnetic field $F_{\mu\nu}$.  Using (\ref{328}), we
can convert the divergence integrals of $\mathbf{E}^i$ and $\mathbf{B}^i$ into
flux integrals of electric and magnetic fields over a surface at infinity and
obtain
\begin{eqnarray}\label{329}
{\cal E} & = & \sin\theta Q_E + \cos\theta Q_M + \frac{1}{2} \int \,d^3x
\partial_i \left( \frac{\sqrt{-{g}}}{g_{00}} \mathbf{\chi}^2 \partial^i f
\right) ,
\end{eqnarray}
where
\begin{eqnarray}\label{330}
Q_E & = & \lim_{R\to\infty} \int_{S_R} \,d\sigma_R^i \frac{1}{v} \left(
\partial_i W_0\right)\chi^2,
\end{eqnarray}
and
\begin{eqnarray}\label{331}
Q_M & = & \lim_{R\to\infty} \frac{1}{2\alpha v^3} \int_{S_R} \,d\sigma_{R i}
\epsilon^{ijk} \mathbf{\chi} \cdot \left[ \partial_j \mathbf{\chi} , \partial_k
\mathbf{\chi} \right] .
\end{eqnarray}
Which leads to
\begin{eqnarray}\label{332}
Q_M & = & \frac{4\pi}{\alpha} n ,
\end{eqnarray}
where $n$, an integer, is the winding number of the mapping $\chi: S_{\infty}^2
\to S^2$,
\begin{eqnarray}
n = \frac{1}{8\pi v^3} \int_{S_{\infty}^2} \,d\sigma_i \epsilon^{ijk}
\mathbf{\chi} \cdot \left[ \partial_j \mathbf{\chi} , \partial_k \mathbf{\chi}
\right] . \nonumber
\end{eqnarray}
$Q_E$ has no such interpretation, but its finiteness leads to a condition on
$W_0$.

If we define
\begin{eqnarray}
\sin\theta = \frac{Q_E}{\sqrt{Q_E^2 + Q_M^2}} & , & \cos\theta =
\frac{Q_M}{\sqrt{Q_E^2 + Q_M^2}} , \nonumber
\end{eqnarray}
then
\begin{eqnarray}\label{333}
{\cal E} \geq \sqrt{Q_E^2 + Q_M^2} + {1\over 2}\int \,d^3x \partial_i \left(
\frac{\sqrt{-{g}}}{g_{00}} \chi^2 \partial^i f \right) ,
\end{eqnarray}
the equality holding when equation (\ref{319}) is satisfied.

\section{Basic Equations with Spherically Symmetric Ansatz}\cleqn
The considerations in the previous section are valid for any compact
group ${\cal G}$.  In this section, we shall specialize to $SU(2)$.  
We define a
spherically symmetric static metric
\begin{eqnarray}\label{41}
ds^2 = A^2(r) dt^2 - B^2(r) dx^2 - r^2\left(d\theta^2 + \sin^2\theta d\varphi^2
\right) ,
\end{eqnarray}
and assume the spherically symmetric 't Hooft-Polyakov 
ansatz~\cite{Polyakov} for the gauge and Higgs fields,
\begin{eqnarray}
A_0^a = \frac{\hat{x}^a}{\alpha} J(r) , & \label{42} \\
\eta_{ab} A_i^b = \epsilon_{aij}\hat{x}^j \frac{\left(1-W(r)\right)}{ \alpha r} ,
& \eta_{ab} = (-1,-1,-1) , \label{43}
\end{eqnarray}
and
\begin{eqnarray}\label{44}
\Phi^a = \hat{x}^a v H(r) .
\end{eqnarray}

In spherical polar coordinates,

\begin{eqnarray}\label{45}
\begin{array}{ll}
A_0^a = \displaystyle{\frac{J(r)}{\alpha}} \left[ 
\begin{array}{c}
\sin\theta\cos\varphi \\
\sin\theta\sin\varphi \\
\cos\theta
\end{array} \right] , \quad 
&   
A_r^a = 0 ,  \\ 
A_{\theta}^a = \displaystyle{\frac{1-W(r)}{\alpha}} \left[ 
\begin{array}{c}
\sin\varphi \\
-\cos\varphi \\
0
\end{array} \right] , \quad &  
A_{\varphi}^a = \displaystyle{\frac{1-W(r)}{\alpha}} \left[ 
\begin{array}{c}
\cos\theta\cos\varphi \\
\cos\theta\sin\varphi \\
-\sin\theta
\end{array} \right] , \quad
\end{array}
\end{eqnarray}
leading to the field strengths
\begin{eqnarray}
\begin{array}{lll}
F_{0r}^a = -\displaystyle{\frac{J'(r)}{\alpha}}\left[ 
\begin{array}{c}
\sin\theta\cos\varphi \\
\sin\theta\sin\varphi \\
\cos\theta
\end{array} \right] , \quad 
&
F_{0\theta}^a = -\displaystyle{\frac{J(r)W(r)}{\alpha}}\left[ 
\begin{array}{c}
\cos\theta\cos\varphi \\
\cos\theta\sin\varphi \\
-\sin\theta
\end{array} \right] , \\
&\\
F_{0\varphi}^a = -\displaystyle{\frac{J(r)W(r)}{\alpha}}\left[ 
\begin{array}{c}
-\sin\varphi \\
\sin\theta\sin\varphi \\
\cos\theta
\end{array} \right] , \quad
&
F_{r\theta}^a = -\displaystyle{\frac{W'(r)}{\alpha}}\left[ 
\begin{array}{c}
\sin\varphi \\
-\cos\varphi \\
0
\end{array} \right] , \\
&\\
F_{r\varphi}^a = -\displaystyle{\frac{W'(r)}{\alpha}}\left[ 
\begin{array}{c}
\cos\theta\cos\varphi \\
\cos\theta\sin\varphi \\
-\sin\theta
\end{array} \right] \sin\theta , \quad 
&
F_{\theta\varphi}^a = \displaystyle{\frac{W^2(r)-1}{\alpha}}\left[ 
\begin{array}{c}
\sin\theta\cos\varphi \\
\sin\theta\sin\varphi \\
\cos\theta
\end{array} \right] \sin\theta , 
\end{array}
\label{47}
\end{eqnarray}
and
\begin{eqnarray}
&\mathbf{D}_r \Phi^a = v H'(r) \left[ 
\begin{array}{c}
\sin\theta\cos\varphi \\
\sin\theta\sin\varphi \\
\cos\theta
\end{array} \right] ,\quad
\mathbf{D}_{\theta} \Phi^a = v W(r)H(r) \left[ 
\begin{array}{c}
\cos\theta\cos\varphi \\
\cos\theta\sin\varphi \\
-\sin\theta
\end{array} \right] , & \nonumber\\
&\mathbf{D}_{\varphi} \Phi^a =- v W(r)H(r) \left[ 
\begin{array}{c}
\sin\varphi \\
-\cos\varphi \\
0
\end{array} \right]
\sin\theta,& \label{48}
\end{eqnarray}
where `prime' denotes derivatives with respect to $r$.

Expressed in spherical polar coordinates, our ansatz (\ref{34}) that leads to the Bogomol'nyi-type equation takes the form
\begin{eqnarray}
\mathbf{D}_r \mathbf{\chi} + \left( f'(r) - \frac{A'(r)}{A(r)} \right)
\mathbf{\chi} & = & -\frac{A(r)B(r)}{r^2\sin\theta} \mathbf{F}_{\theta\varphi} ,
\label{49} \\
\mathbf{D}_{\theta} \mathbf{\chi} & = & -\frac{A(r)}{B(r)}\frac{1}{\sin\theta}
\mathbf{F}_{\varphi r} , \label{410} \\
\mathbf{D}_{\varphi} \mathbf{\chi} & = & -\frac{A(r)}{B(r)} \sin\theta
\mathbf{F}_{r \theta} . \label{411}
\end{eqnarray}
Substituting the spherically symmetric ansatz for $\mathbf{\chi}$ and
$\mathbf{F}$, we obtain two independent equations,
\begin{eqnarray}
H'(r) + f'(r) H(r) & = & \frac{B(r)\left( 1 - W^2(r) \right)}{\alpha v r^2} ,
\label{412} \\
W'(r) & = & -\alpha v W(r)H(r) B(r) . \label{413}
\end{eqnarray}

The Yang-Mills field equations (\ref{31}) and (\ref{36}), and the Higgs
field equations (\ref{38}) are transformed into
\begin{eqnarray}
J''(r) + \left( \frac{2}{r} - \frac{A'(r)}{A(r)} -
\frac{B'(r)}{B(r)} \right) J'(r) - \frac{2 B^2(r)}{r^2} J(r)W(r)^2
& = & 0 , \hskip 1cm
\label{414} \\
J^2(r)W(r) - \left( f'(r) - \frac{A'(r)}{A(r)} \right) \frac{A^2(r)}{B^2(r)}
W'(r) & = & 0 , \hskip 1cm \label{415} \\
\left( - f'(r) + \frac{A'(r)}{A(r)} \right) \frac{ H'(r)}{H(r)B^2(r)}
+ \frac{1}{B^2(r)} \left( \frac{B'(r)}{B(r)} - \frac{2}{r} \right)
f'(r) - \frac{1}{B^2(r)} f''(r) & = &
\nonumber\\
&& \hskip -4cm\frac{R-2\Lambda}{8\pi Gv^2} + \lambda v^2 \left( H^2(r) - 1
\right).  \label{416}
\end{eqnarray}

We would like to remark that in the case of pure monopole in the
$\mathbf{A}_0 = 0$ gauge then $J(r)=0$.  Thus if 
$f'(r) = \frac{A'(r)}{A(r)}$, the 
Yang-Mills equations (\ref{414}) and (\ref{415}) are 
automatically satisfied and the Higgs field equation (\ref{416}) becomes 
\begin{eqnarray}\label{419}
\frac{1}{B^2(r)} \left[
\left( \frac{B'(r)}{B(r)} - \frac{2}{r} \right)
\frac{A'(r)}{A(r)}-
\left(\frac{A'(r)}{A(r)}\right)'\right]=
\frac{R-2\Lambda}{8\pi Gv^2} + \lambda v^2 \left( H^2(r) - 1
\right) .
\end{eqnarray}

From the metric (\ref{41}), we calculate the components of the Einstein
tensor, $G_{\mu\nu}$, and the Ricci scalar, $R$, and find them to be
\begin{eqnarray}\label{E49}
G_{00} & = & \frac{A^2}{B^2} \left[\frac{1}{r^2}\left( B^2 - 1 \right) +
\frac{2}{r} \frac{B'}{B} \right] , \nonumber \\
G_{rr} & = & \frac{1}{r^2}\left( 1 - B^2  \right) + \frac{2}{r} \frac{A'}{A} ,
\nonumber \\
G_{\theta\theta} & = & \frac{r^2}{B^2} \left( \frac{A''}{A} + \frac{1}{r}
\frac{A'}{A} - \frac{A'}{A} \frac{B'}{B} - \frac{1}{r} \frac{B'}{B} \right) ,
\nonumber \\
G_{\varphi\varphi} & = & \sin^2\theta G_{\theta\theta} ,
\end{eqnarray}
and
\begin{eqnarray}\label{E410}
R = \frac{2}{r^2}\left( \frac{1}{B^2} - 1 \right) + \frac{2}{B^2} \left(
\frac{A''}{A} + \frac{2}{r} \frac{A'}{A} - \frac{A'}{A} \frac{B'}{B} -
\frac{2}{r} \frac{B'}{B} \right) .
\end{eqnarray}
The rest of the components are zero.  Likewise, we express the components of the
energy-momentum tensor using (\ref{49})-(\ref{411}), and find
\begin{eqnarray}\label{421}
T_{00} & = &
\frac{1}{{\alpha}^2}\left[\frac{1}{2B^2}J'^2
+\frac{1}{r^2}J^2W^2+\frac{A^2}{B^2r^2}W'^2
+\frac{1}{2}\frac{A^2}{r^4}\left(W^2-1\right)^2\right]\nonumber\\
&&\hskip 2cm
+v^2\left[\frac{1}{2}\frac{A^2}{B^2}H'^2+\frac{A^2}{r^2}W^2H^2
+\frac{\lambda}{4}A^2v^2\left(H^2-1\right)^2\right],\\
T_{rr} & = &
\frac{1}{\alpha^2}\left[-\frac{1}{2A^2}J'^2
+\frac{B^2}{A^2}J^2W^2+\frac{1}{r^2}W'^2
-\frac{B^2}{2r^4}\left(W^2-1\right)^4\right]\nonumber\\
&&\hskip 2cm 
+v^2\left[\frac{1}{2}H'^2-\frac{B^2}{r^2}W^2H^2
-\frac{\lambda}{4}B^2v^2\left(H^2-1\right)^2\right],\\
T_{\theta\theta} & = &
\frac{1}{2\alpha^2}\left[\frac{1}{A^2B^2}r^2J'^2
+\frac{1}{r^2}\left(W^2-1\right)^2\right]
-\frac{v^2}{2}\left[\frac{r^2}{B^2}H'^2
+\frac{\lambda}{2}r^2v^2\left(H^2-1\right)^2\right],\\
T_{\varphi\varphi}& = & \sin^2\theta T_{\theta\theta}.
\end{eqnarray}
Using (\ref{E49}) and (\ref{421}), it is straight forward to write Einstein field
equations
\begin{eqnarray}
G_{\mu\nu} + \Lambda g_{\mu\nu} = \frac{8\pi Gv^2}{\mathbf{\Phi}^2} 
T_{\mu\nu} .
\nonumber
\end{eqnarray}
In what follows, we eliminate from them their dependence on $W'(r)$ 
and $W(r)$ by using equations (\ref{412}) and (\ref{413}) and 
take linear combinations of the resulting equations. These are
\begin{eqnarray}
\frac{1}{r}\left(\frac{A'}{A}+\frac{B'}{B}\right) & = & 8\pi G
\left(\frac{v^2}{2}\left(\frac{H'}{H}\right)^2
+B^2\left(1-v\alpha r^2\left({H'+f'H\over B}
\right)\right)\left(\frac{1}{\alpha^2 r^2}\frac{J^2}{A^2H^2}+
\frac{v^2}{r^2}\right)\right)\label{422},
\end{eqnarray}
\begin{eqnarray}
\frac{A''}{A}-\frac{A'}{A}\frac{B'}{B}+\frac{B^2-1}{r^2} & = &
8\pi G\left[\frac{1}{\alpha^2}\frac{J'^2}{A^2H^2} +
\frac{1}{2}v^2\left(\frac{H'}{H}\right)^2+\frac{v^2B^2}{r^2}
\right.\nonumber\\
&& \left.\hskip 2cm +f'v^2\left(2\frac{H'}{H}+f'\right)
-v^3B\alpha\left(H'+f'H\right)\right], \label{423}
\end{eqnarray}
\begin{eqnarray}
\hskip -.2cm
\frac{A''}{A}+\frac{1}{r}\left(\frac{A'}{A}-\frac{B'}{B}\right)
-\frac{A'}{A}\frac{B'}{B}+
\Lambda B^2 \hskip -.2cm&=& \hskip -.2cm
4\pi G\left[ \frac{1}{\alpha^2}\frac{J'^2}{A^2H^2}
+f'v^2\left(2\frac{H'}{H}+f'\right) 
-\frac{1}{2}B^2v^4\frac{\left(H^2-1\right)^2}{H^2}\right].
\label{424}
\end{eqnarray}

\section{Higgs Vacuum; Monopole and Dyon Solutions}
\cleqn
We seek solutions to the set of coupled non-linear equations 
(\ref{412})-(\ref{416}) and (\ref{422})-(\ref{424}) in the
Higgs vacuum, that is, the Higgs field frozen to a constant, $H=1$. 
The equations simplify considerably.
To this end, we introduce the dimensionless
co-ordinate $x$ and define dimensionless $J$,
\begin{eqnarray}
x=\alpha vr,\qquad  \frac{J}{\alpha v}\rightarrow J \label{51}.
\end{eqnarray}

We shall also work  in units of the dimensionless coupling $4\pi Gv^2$ set equal to unity and
henceforth all the field variables are functions of $x$ and `prime' will denote derivatives with respect to $x$.
We shall suppress their dependence on $x$ for simplicity.

From eqns (\ref{413}) and (\ref{415}), it follows that
\begin{eqnarray}
f'&=& \frac{A'}{A}-\frac{J^2B}{A^2} \label{52},
\end{eqnarray}
and hence the Bogomol'nyi conditions (\ref{412}) and (\ref{413}) take
the form
\begin{eqnarray}
\frac{A'}{AB}& = & \frac{J^2}{A^2}+\frac{B(1-y)}{x^2}\label{53}, \\
B & = & -\frac{1}{2}\frac{y'}{y}\label{54},
\end{eqnarray}
where we have defined $y$=$W^2$.

Then the remaining Yang-Mills, Higgs and Einstein field equations 
(\ref{414}), (\ref{416}), (\ref{422})-(\ref{424}) are given by
\begin{eqnarray}
J''+\left(\frac{2}{x}-\frac{A'}{A}-\frac{B'}{B}\right)J'
-\frac{2}{x^2}B^2Jy&=&0\label{55},
\end{eqnarray}
\begin{eqnarray}
B^2\Lambda+{B^2-1 \over x^2}
-2\frac{A''}{A}+\frac{2}{x}\frac{B'}{B}
+\frac{A'}{A}\left(\frac{A'}{A}+2\frac{B'}{B}-\frac{4}{x}\right)
&=& 2J\frac{B}{A^2}\left(J\left(\frac{A'}{A}-\frac{1}{x}\right)
-J'\right)\label{56},
\end{eqnarray}
\begin{eqnarray}
\frac{1}{x}\left(\frac{A'}{A}+\frac{B'}{B}\right)-\frac{2B^2}{x^2}
\left(1-x^2\frac{A'}{AB}\right) &=&
2J^2\frac{B^2}{A^2}\left(1+\frac{1}{x^2}-\frac{A'}{AB}
+\frac{J^2}{A^2}\right) ,  \label{57}
\end{eqnarray}
\begin{eqnarray}
\frac{A''}{A}-\frac{A'}{A}\left(\frac{B'}{B}+2\frac{A'}{A}-2B\right)
-{B^2+1 \over x^2}&=&
2\frac{J'^2}{A^2}+2J^2\frac{B^2}{A^2}\left(1-{2\over B}\frac{A'}{A}+
\frac{J^2}{A^2}\right) , \label{58}
\end{eqnarray}
\begin{eqnarray}
\frac{A''}{A}+\frac{1}{x}\left(\frac{A'}{A}-\frac{B'}{B}\right)
-\frac{A'}{A}\left(\frac{A'}{A}+\frac{B'}{B}\right) 
+ \Lambda B^2&=&
\frac{J'^2}{A^2}-2B\frac{J^2}{A^2}\frac{A'}{A}+B^2\frac{J^4}{A^4} .
\label{59}
\end{eqnarray}

\subsection{Monopole solutions}

Equations for the monopole are obtained by setting $J=0$ in the 
above set of equations. At the outset, we note
that eqn. (\ref{55}) is automatically satisfied. By taking suitable 
linear combinations of (\ref{56}), (\ref{57}), (\ref{58}) 
and (\ref{59}), we find
\begin{eqnarray}
\Lambda\left(3B^2+1\right)=0,
\end{eqnarray}
which implies $\Lambda=0$ for $B$ to be real.  Thus, there is no
monopole solution for non-vanishing cosmological constant in our
model. With $\Lambda=0$, we are left with three independent
equations, 
\begin{eqnarray}
\frac{A'}{A}=\frac{1}{x^2}B\left(1-y\right),\quad
 B=-\frac{1}{2}\frac{y'}{y},\quad
 B=1+x\frac{A'}{A}\label{512},
\end{eqnarray}

or equivalently,
\begin{equation}
y'=-\frac{2xy}{x+y-1}, \label{513} 
\end{equation}
\begin{equation}
B =\frac{x}{x+y-1},\quad
\frac{A'}{A}=\frac{1-y}{x\left(x+y-1\right)}\label{515}.
\end{equation}

Equation (\ref{513}) is the well known Abel's differential equation of 
the second type. The two equations in (\ref{515}) are 
determined in terms of the solutions of Abel's equation.
Abel's equation has no known analytical solution other than
$y=0$, in which case, we have an Abelian magnetic monopole with
metric functions given by

\begin{eqnarray}
A^2=\left(1-\frac{1}{x}\right)^2,\qquad B^2=
{1\over A^2}=\left(1-\frac{1}{x}\right)^{-2}\label{516}.
\end{eqnarray}

This Abelian monopole solution is also an extreme 
Reissner-N\"{o}rdstrom black hole with mass $M$ and charge
$Q$ given by

\begin{eqnarray}
M=\frac{4\pi Gv}{\alpha},\qquad Q=\frac{M}{\alpha v}\label{517}.
\end{eqnarray}

\begin{figure}[ht]
\centerline{\epsfig{file=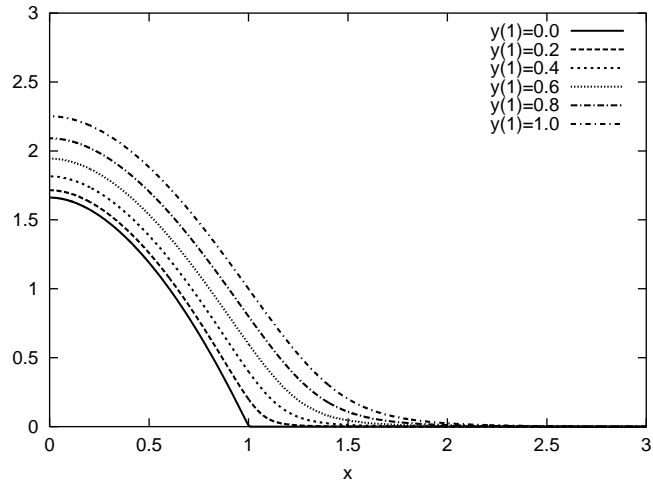,width=3.5in}}
\caption[]{Plot of family of solutions to Abel's equation, i.e. $y(x)$, for 
a range of initial conditions $y(1)=0,0.2,0.4,0.6,0.8,1.0$.} 
\label{solutions}
\end{figure}

\begin{figure}[ht]
\centerline{\epsfig{file=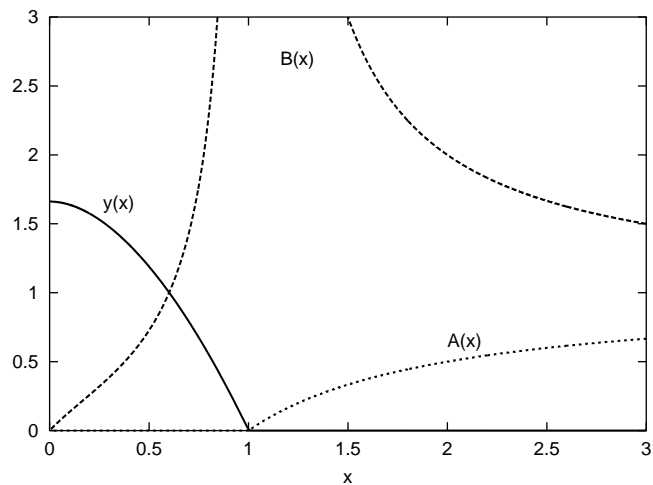,width=3.5in}}
\caption[]{Plot of $A(x)$, $B(x)$ and $y(x)$ against $x$ for an 
extremal non-Abelian monopole ($y(1)=0$). This solution is a 
black hole since we see the event horizon at $x=1$.}
\label{extremalmonopole}
\end{figure}

\begin{figure}[ht]
\centerline{\epsfig{file=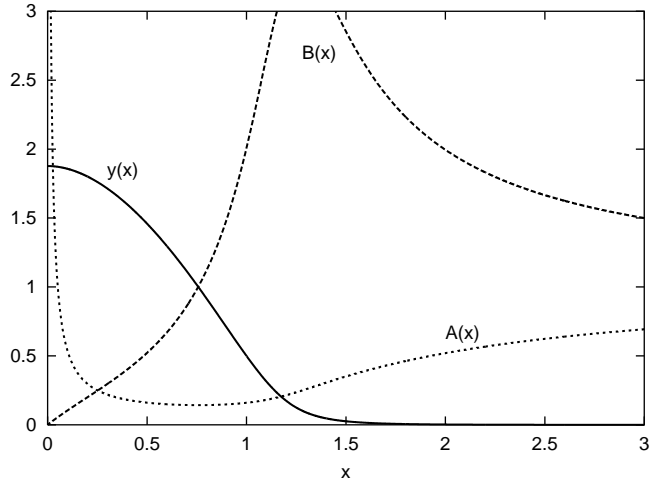,width=3.5in}}
\caption[]{Plot of $A(x)$, $B(x)$ and $y(x)$ against $x$ for a non-Abelian monopole with the initial condition $y(1)=0.5$.}
\label{nonextremalmonopole}
\end{figure}

In the general case, with non-vanishing $y$, we can solve equation  
(\ref{513}) numerically. Results are shown in Figs~
\ref{solutions}, \ref{extremalmonopole} and \ref{nonextremalmonopole}.

We observe that we have a family of solutions that have exponentially
vanishing non-Abelian components $y=W^2$. They are 
determined by the initial condition $y(1)=y_{\rm init}$ 
(see Fig.~\ref{solutions}). These solutions are characterized by
a non-Abelian core, outside of which only an Abelian component 
remains. The solutions to the metric components $A$ and $B$ in
general have a minimum and maximum respectively in the vicinity of $x=1$. From
eqn.~(\ref{515}) the metric component $A$ is dependent on an 
initial condition, however, once one fixes its asymptotic behavior 
($A\rightarrow 1$ as $x\rightarrow\infty$) it is uniquely determined.
The metric  component $B$ automatically has the correct asymptotic
behavior. We see that as the initial value of $y$ at $x=1$ 
approaches zero, the minimum of $A$ also approaches zero at $x=1$ 
(and the maximum $B$ at $x=1$ gets larger). The solution for $y(1)=0$ 
(see Fig.~\ref{extremalmonopole}) represents the extremal case 
where an event horizon has formed. This extremal solution is a 
black hole with a non-Abelian magnetic monopole.

For the extremal solution, the non-Abelian magnetic field is
confined within the black hole horizon. Thus it is only natural that the
metric coefficients outside the horizon are identical to that of
a Reissner-N\"{o}rdstrom black hole. In Fig.~\ref{extremalmonopole}
we see that the metric coefficient for $A$ vanishes inside the 
horizon, $x\leq1$. This is not a problem since $A$ has been normalized 
at $x=\infty$. An observer at infinity can not observe the interior 
of the black hole since for him an object takes an infinite amount
of time to reach the horizon. If one choses to normalize $A$ at the
origin one would have a perfectly well defined metric inside 
the horizon which is infinite outside. This does not affect the 
determination of the non-Abelian magnetic field $y$ which is
independent of the normalization of $A$. This behaviour is also
observed in the monopole solutions of Lue and Weinberg~\cite{Lue}.

In Fig.~\ref{nonextremalmonopole} we see a non-extremal monopole
solution that is not a black hole.

\subsection{Dyonic solutions}

The set of seven coupled non-linear equations (\ref{53})-(\ref{59}) involve
four functions, $y$, $J$, $A$ and $B$. 
We do not expect to find analytic solutions to this set of equations. 
Thus we are forced to look for numerical solutions.

First, however we can analyse some features of these equations.
Equation (\ref{57}) is a quadratic equation in $J^2/A^2$ which has
the solutions
\begin{equation}
2\frac{J^2}{A^2}=-\left(1+\frac{1}{x^2}-\frac{A'}{AB}\right)\pm\sqrt{\left(1-\frac{1}{x^2}
+\frac{A'}{AB}\right)^2+\frac{2}{xB^2}\left(\frac{A'}{A}+\frac{B'}{B}\right)}\label{518}.
\end{equation}

Combining equations (\ref{53}) and (\ref{518}) for $J^2/A^2$, we obtain
\begin{equation}
\left(1-\frac{1}{x^2}+\frac{A'}{AB}\right)+
\frac{y}{x^2}=\pm\sqrt{\left(1-\frac{1}{x^2}
+\frac{A'}{AB}\right)^2+\frac{2}{xB^2}\left(\frac{A'}{A}+\frac{B'}{B}\right)} .
\end{equation}

By inspection, we see that if $y=0$, the case of an Abelian dyon, then the above equation implies that 
\begin{equation}
\frac{A'}{A}+\frac{B'}{B}=0,
\end{equation}
assuming $A,B \neq 0$. This in turn implies that the product 
$AB$ is constant.

Thus the Yang-Mills field equation (\ref{55}) reduces to
\begin{eqnarray}
J''+2J'=0,
\end{eqnarray}
which has the analytic solution
\begin{eqnarray}
J(x)=c_1+c_2/x,
\end{eqnarray}
where $c_1,c_2$ are arbitrary constants.

Furthermore, if the charge distribution is assumed to vanish 
asymptotically, then $c_1=0$. Together with the
asymptotic conditions on the metric functions $A$, $B$, we can show
that the cosmological constant $\Lambda$ must vanish. However, 
inspection of equation~(\ref{59}) leads to a contradiction; 
which proves that there are no Abelian dyonic solutions to our 
equations (in contrast to the monopole case). 

For the general case of non-Abelian dyon solutions with $y\neq0$, 
examination of the asymptotic behaviour of equation~(\ref{59}) 
reveals that for consistent solutions the cosmological constant must
vanish. Proceeding, we find two coupled equations 
in $y$ and $J$, whose solutions yield consistent solutions to all 
the equations. These are
\begin{eqnarray}
y''yx^3-\frac{1}{2}y'^3\left(x^2-2y\right)
-\frac{1}{2}y'^2\left(1-7y+2x^3\right)+2y'yx^2 
+2yx^2\left(x+y'\right)\left(y'+y\frac{J'}{J}\right)\frac{J'}{J}=0
\label{2ndorder},
\end{eqnarray}
\begin{eqnarray}
&&\left(\frac{J'}{J}\right)^4+\frac{y'}{y}\left(\frac{J'}{J}\right)^3-
\left(\frac{1}{x}+\frac{y'}{2x^2}\right)\left(\frac{y'}{y}\right)^2
\left(\frac{J'}{J}\right) \nonumber\\
&& \hskip 2cm
+{1\over x^2}
\left(\frac{y'}{y}\right)^2\left[\left(\frac{y'}{y}\right)^2{\left(x^2-5y^2+2y-1\right)\over16x^2}
+\frac{y'}{y}{\left(1-5y\right)\over4x}-{5\over4}\right]=0\label{quartic}.
\end{eqnarray}

The solutions to $y$ and $J$ yield the metric functions $A$ and $B$ via
the following equations:
\begin{eqnarray}
{A'\over A} &=& -{1\over 2}{y'\over y}\left(1+{1-3y \over x^2}\right)  
+2{y\over y'} \left({J'\over J}+{1\over 2}{y'\over y}\right)^2
+{2\over x} , \\
B &=& -{1\over 2}{y'\over y} .
\end{eqnarray} 

Equation~(\ref{quartic}) is a quartic polynomial in $J'/J$ in terms 
of $y$ and $y'$. Since $J$ is positive and the requirement that $J$  
vanish asymptotically implies that $J'/J$ should be negative 
everywhere (as long as $J$ is montonic). Thus we select the 
negative real root of the quartic polynomial.   
Equation~(\ref{2ndorder}) is a non-linear second order differential
equation in $y$ which can be readily integrated numerically using a
4th order Runga-Kutta with initial conditions set at $x=1$. 

Inspection of equations~(\ref{2ndorder}) and (\ref{quartic}) reveals that 
in order for $y$ to vanish, it can only occur at $x=1$.  
One can perform a similar inspection of Abel's equation in the monopole
case where it is also evident that $y$ can also only vanish at $x=1$. 
Thus, like in the monopole case one expects that the extremal black hole 
solution will be reach as the initial condition for $y(x=1)\rightarrow 0$. 
However, unlike the monopole case there is an additional parameter, the
initial value of $y'$ at $x=1$. 

\begin{figure}[ht]
\centerline{\epsfig{file=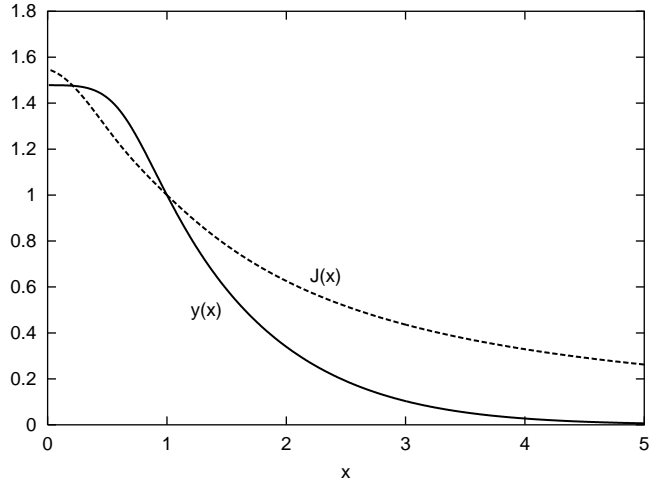,width=3.5in}}
\caption[]{Plot of $y(x)$ and $J(x)$ against $x$ for a non-Abelian 
dyon with initial conditions $y(1)=1$, $y'(1)=-1$, $A(1)=2$, $J(1)=1$.} 
\label{yandJ1}
\end{figure}
\begin{figure}[ht]
\centerline{\epsfig{file=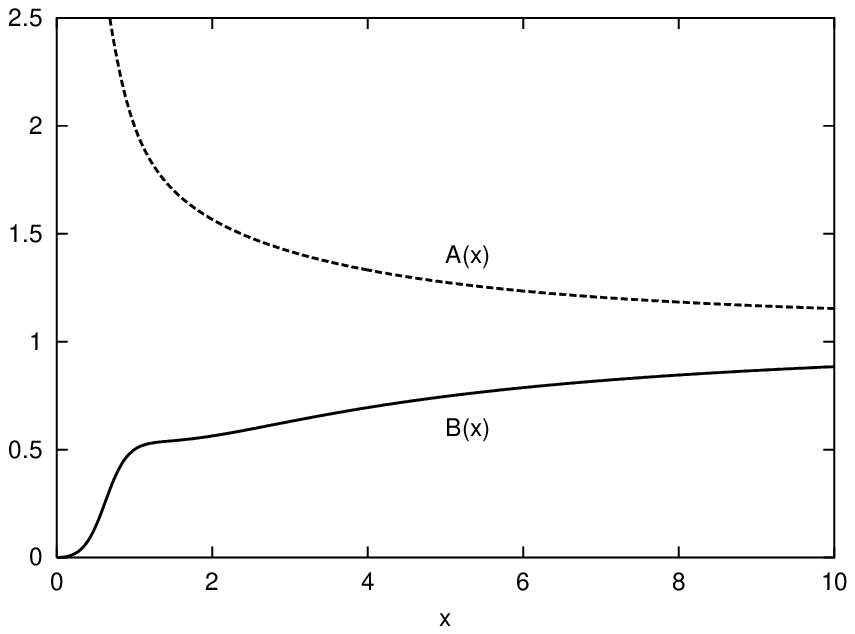,width=3.5in}}
\caption[]{Plot of $A(x)$ and $B(x)$ against $x$ for a non-Abelian 
dyon with initial conditions $y(1)=1$, $y'(1)=-1$, $A(1)=2$, $J(1)=1$.} 
\label{AandB1}
\end{figure}
\begin{figure}[ht]
\centerline{\epsfig{file=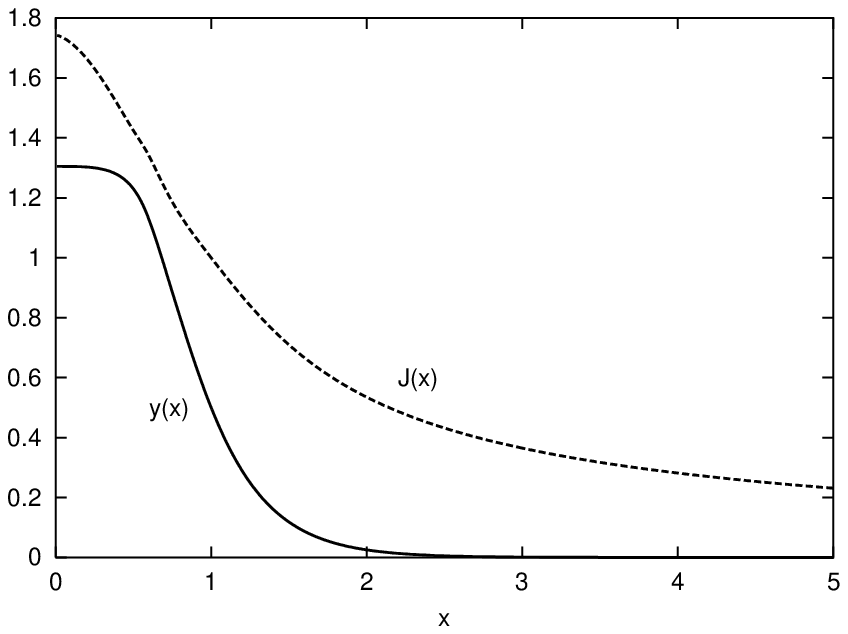,width=3.5in}}
\caption[]{Plot of $y(x)$ and $J(x)$ against $x$ for a non-Abelian 
dyon with initial conditions $y(1)=0.5$, $y'(1)=-1.3$, 
$A(1)=0.4$, $J(1)=1$.} 
\label{yandJ2}
\end{figure}
\begin{figure}[ht]
\centerline{\epsfig{file=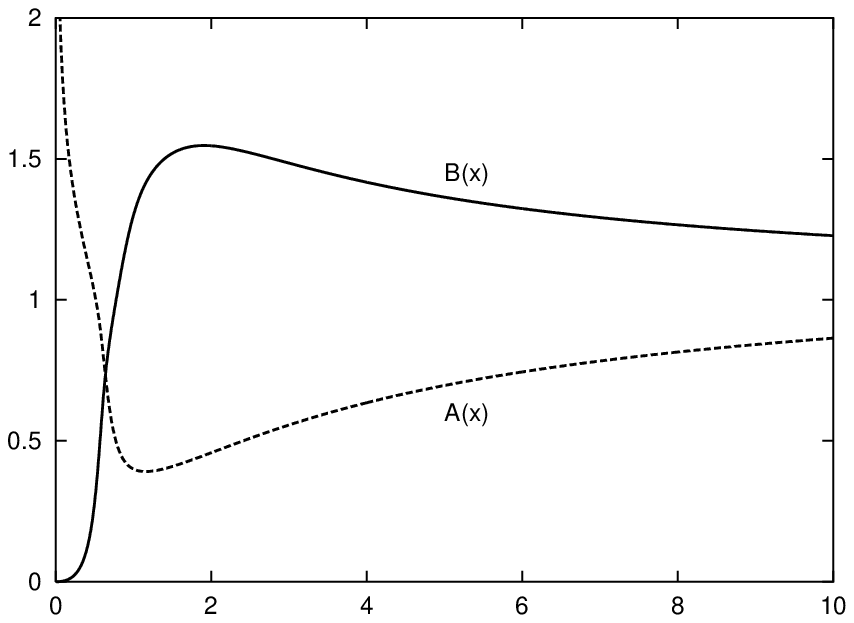,width=3.5in}}
\caption[]{Plot of $A(x)$ and $B(x)$ against $x$ for a non-Abelian 
dyon with initial conditions $y(1)=0.5$, $y'(1)=-1.3$, 
$A(1)=0.4$, $J(1)=1$.} 
\label{AandB2}
\end{figure}
Figures~\ref{yandJ1} and \ref{AandB1} contain plots of the solutions for
$y$, $J$, $A$ and $B$ for initial conditions $y(1)=1$, $y'(1)=-1$,
$A(1)=2$ and $J(1)=1$. The initial value of $A$ is set in order to obtain 
it's correct asymptotic behavior. For these initial conditions the metric
function $A$ does not have a minimum and $B$ does not have a maximum 
at finite $x$.
However, as one adjusts the initial conditions closer to the critical
value, $y(1)=0$, a minima and maxima begin to appear. 
This is evident in figures~\ref{yandJ2} and \ref{AandB2} where the 
initial conditions are $y(1)=0.5$, $y'(1)=-1.3$, $A(1)=0.4$, $J(1)=1$.

While inspecting the parameter space of initial conditions for $y'(1)$
one finds that $y'(1) \in (-2,0)$ in order for solutions to exist.
For a value of $y'(1)$ in this range one finds that $y(1)$ can not be
made arbitrarily close to zero if one requires the numerical solution
to be continuously defined for all $x$. The smallest value of $y(1)$ 
for a well defined solution depends on the value of $y'(1)$. 
Solutions for values of $y(1)$ closer to zero exhibit numerical 
singularities in the region $x<1$. 
The probable explanation for this is that unlike an extremal 
monopole which only has only one horizon, a dyonic black hole
naturally has two horizons. Thus, as the solution approaches the
black hole limit the appearance of one of the horizons might occur first 
(meaning the metric coefficient $A$ becomes small and $B$ becomes 
large at some value of $x$). Also, with the existence of two horizons 
it is clear that our
metric (equation~(\ref{41})), which has positive coefficients $A^2$ \
and $B^2$, can not properly describe the region between two 
horizons where the space-like and time-like co-ordinates flip.
Thus, it is not unreasonable that as our dyonic solution becomes a
black hole (by tuning the initial conditions closer to their critical 
values), that the co-ordinates we used are not appropriate for all values
of $x$.

\section{Conclusions}

We have shown in Section 3 that, with an ansatz, eqn.~(\ref{34}), 
we are led to Bogomol'nyi type equations and an energy functional that 
is bounded from below in the case of a dyon. In flat space-time, 
solutions that saturate the Bogomol'nyi bound are known to exist 
\cite{Coleman}. Demonstrating this in the case of a non-Abelian dyon 
in curved space-time is a new result. These results follow from 
the form of the action in eqn.~(\ref{21}), where the Higgs 
scalar is directly coupled 
to the Ricci scalar. In the monopole case, this unconventional 
coupling led to a first order classic Abel's equation, the 
solution of which yielded solutions to all the equations in 
the problem \cite{Nguyen}. In the dyon case, the problem turns out to be 
more complicated; nonetheless, contains similar simplifications 
in contrast to the more standard approaches in the literature.

In Section 5, we specialize to Higgs vacuum solutions to the general 
equations in Section 4. For the sake of completeness, we begin with 
a review of the non-Abelian monopole solutions \cite{Nguyen}. We expand 
on the discussions therein. These monopole solutions depend upon 
initial conditions; for a specific choice the solution represents 
a non-Abelian extremal black hole. In addition we show that in our
model there are no solutions with a non-vanishing cosmological
constant in contrast to the conventional model in \cite{Brihay}.
We also present numerical evidence 
for non-Abelian dyonic solutions. For a range of initial conditions 
the defining equations yield stable numerical results which are 
well-defined to the center of the dyon. However, as the initial 
conditions are tuned closer to critical values, the numerical 
results have singularities that likely correspond to the formation 
of horizons. Also as in the monopole case, physically interesting
solutions require a vanishing cosmological constant, and further,
unlike in the case of the monopole, there is no Abelian dyon 
solution in our model.          

Finally, to the best of our knowledge, a direct proof of Dirac charge 
quantization  in the case of the EYMH system does not exist. However, 
there are compelling reasons to think, that the topological 
considerations in the case of flat space-time can be extended to 
curved space-time. If so, it has some extremely interesting 
consequences that have been pointed out by Ignatev, Joshi and Wali~
\cite{Ignatev}. Among them, are the consequences that magnetically 
or electrically charged black holes obey a lower bound on their 
mass and the consequence that is not widely appreciated, namely, 
the spectrum of magnetically (or electrically) charged extremal 
black holes are evenly spaced in mass due to charge quantization.    

\underline{Acknowledgments:} 
This work was supported in part (JSR and KCW) by the U.S Department 
of Energy (DOE) under contract no. DE-FG02-85ER40237. It was also 
supported in part (KCW) by a grant from the NSF, Division Of International 
Programs (U.S. - Australia Cooperative Research), no. INT-9802698. 
JSR and KCW would like to thank R. F. Sawyer for his invaluable 
help and suggestions in finding numerical solutions and 
also P. Silva for many helpful discussions.

\end{document}